\documentclass[sn-mathphys]{sn-jnl}% Math and Physical Sciences Reference Style
\jyear{2022}%

\theoremstyle{thmstyleone}%
\theoremstyle{thmstyletwo}%

\theoremstyle{thmstylethree}%

\usepackage{graphicx}
\usepackage[flushleft]{threeparttable}
\usepackage{float}
\usepackage{tablefootnote}

\usepackage[compact]{titlesec}        
\titlespacing{\section}{8pt}{8pt}{8pt} 
\usepackage{hyperref}
\begin{document}

\title[Graph Neural Networks for ECG Classification]{Graph Neural Networks for Topological Feature Extraction in ECG Classification}


\author*[1]{\fnm{Kamyar} \sur{Zeinalipour}}\email{kamyar.zeinalipour2@unisi.it}

\author[1,2]{\fnm{Marco} \sur{Gori}}\email{marco.gori@unisi.it}

\affil*[1]{\orgdiv{Department of Information engineering and mathematics}, \orgname{University of Siena}, \orgaddress{\street{Via Roma, 56}, \city{Siena}, \postcode{53100}, \country{Italy}}}

\affil[2]{\orgdiv{3IA C\^ote d'Azur}, \orgname{Universit\'e C\^ote d'Azur}, \orgaddress{\street{28 Avenue de Valrose}, \city{Nice}, \postcode{06000}, \country{France}}}


\abstract{The electrocardiogram (ECG) is a dependable instrument for assessing the function of the cardiovascular system. There has recently been much emphasis on precisely classifying ECGs. While ECG situations have numerous similarities, little attention has been paid to categorizing ECGs using graph neural networks. In this study, we offer three distinct techniques for classifying heartbeats using deep graph neural networks to classify the ECG signals accurately. We suggest using different methods to extract topological features from the ECG signal and then using a branch of the graph neural network named graph isomorphism network for classifying the ECGs. On the PTB Diagnostics data set, we tested the three proposed techniques. According to the findings, the three proposed techniques are capable of making arrhythmia classification predictions with the accuracy of 99.38, 98.76, and 91.93 percent, respectively.}




\maketitle

\section{Introduction}

One of the biophysical signals that may be monitored using special equipment from the human body is electrocardiography (ECG). It stores crucial information about how the heart functions and whether it is affected by aberrant conditions. Cardiologists and medical practitioners frequently utilize ECG to check heart health. The difficulty in recognizing and classifying diverse waveforms and morphologies in ECG signals, like with many other time-series data, is the fundamental issue with manual analysis. This task would take a human a long time to complete and is prone to mistakes. It's worth noting that accurate detection of cardiovascular disorders is critical because they account for around one-third of all fatalities worldwide. Millions of individuals suffer from irregular heartbeats, which can be fatal in some situations. As a result, a precise and low-cost arrhythmic heartbeat diagnosis is exceptionally desirable.\\
Many studies in the literature investigated utilizing machine learning approaches to reliably detect abnormalities in ECG data to overcome the challenges identified by human analysis. Most of these methods include a signal preparation step. Afterward, handmade features are derived from these signals, typically statistical summarizations of signal windows, and employed in subsequent analysis for the final classification task. Powerful and reliable algorithms have been used to divide ECG signals into two or more classes in computer-based applications, including fuzzy c-means~\cite{yeh2010novel}. k-nearest neighbor~\cite{sharma2018novel}, deep neural network~\cite{al2016deep}, convolution neural network~\cite{hammad2019novel,diker2019novel}, higher-order statistics, and the Naive–Bayes classifier~\cite{marinho2019novel}, support vector machines~\cite{mondejar2019heartbeat}, and Weighted ELM~\cite{utomo2019automatic}.\\
The main contributions of this study are as follows:
The first contribution is that we propose to combine different methods to develop a novel approach for ECG classification. To accomplish this, we transform ECG signals into graphs using three distinct ways, then feed the resulting graph to a graph neural network branch called Graph Isomorphism Network (GIN) for graph classification. The second contribution is that we performed an experimental comparison on a well-known data set (PTB Diagnostic ECG~\cite{goldberger2000physiobank} and~\cite{bousseljot1995nutzung}), which demonstrated the capability of the proposed approaches. The third contribution is that we shared public code\footnote{https://pypi.org/project/ts2g, https://github.com/KamyarZeinalipour/ts2g} for transforming time series into graphs using different methods that we will discuss in this study; therefore that other researchers can follow in our footsteps.\\
The remainder of this paper is arranged as follows: the proposed methodology is presented in the second section, experiments of the suggested technique on the ECG classification task are shown in section three, along with a comparison to comparable work in the literature, and finally, section four by a conclusion brings the paper to a close.

%------------------------------------------------

\section{Methodology}

To classify the ECG signal, we utilized three types of transform methods in order to transform time series into the graphs; these methods include natural visibility graph~\cite{lacasa2008time}, horizontal visibility graph~\cite{luque2009horizontal}, and quantile graph~\cite{campanharo2011duality} then the corresponded graphs feed into the graph isomorphism network to create a model for classifying the ECGs, This is the entire procedure we used in this investigation.

\subsection{From time series to complex network}
In this part, we describe key results achieved in the literature on the time series mapping approaches to a complex network that we implemented in our study. These mapping methods are based on visibility concepts and probabilities of transition.

\subsubsection{Natural visibility graph}
Lacasa et al. initially developed the first solution for transforming time series into graphs based on the visibility principle in 2008~\cite{lacasa2008time}. Natural Visibility Graph (NVG) is the name given to this approach, each node in the graph corresponds to the time series data in the same sequence, and two nodes are connected if a line of visibility exists between the respective data points, i.e. if in the time series a straight line is possible to be drawn between the two corresponding data points that do not intercept the data "height" between the two data points of the time series. This approach is built on the assumption that each time series data is perceived as a vertical bar with a height equal to the corresponding numerical value and that the top part of the bar is visible from the top part of the other bars when we examine these vertical bars in the landscape, where we regard each time data as a node in a graph, two nodes are linked if the tops of the respective vertical bar are visible to another, that is, if the tops of the two bars have a straight line that does not cross other bars. The mapping of a toy time series and the resultant network is demonstrated in the left side Figure~\ref{fig:1}, each data in the series corresponds to a network node, and two nodes are connected if their associated data heights satisfy equation 1 visibility condition, the blue lines in the time series plot show the visibility lines (and consequently the connections in the graph) on the other hand as shown in the Figure the generated graph contains as many vertices as the amount of time series data (observation). Formally, each vertex is an observation  $(t_a,y_a )$ and two nodes  $(t_a,y_a )$ and $(t_b,y_b )$  are linked  (with visibility), If any additional observation  $(t_c,y_c )$  with $t_a<t_c<t_b$ fulfills:\\
\begin{equation}  y_c<y_b+(y_a-y_b)\frac{t_b-t_c}{t_b-t_a}   \end{equation}

\subsubsection{Horizontal visibility graph}
\begin{figure}
    \begin{center}
       \includegraphics[width = 1\textwidth]{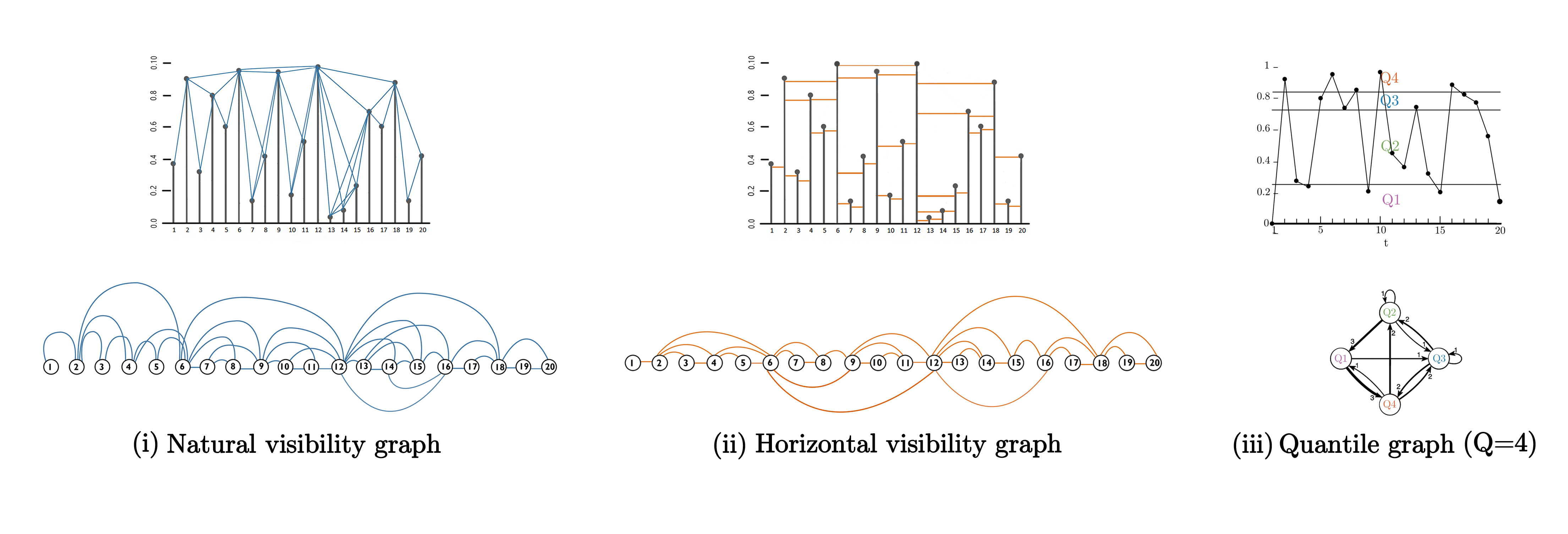} 
    \end{center}
    \caption{
This Figure includes three time series and their corresponding graphs, demonstrating different techniques for converting time series to the graph. Toy time series are depicted in the upper half of the picture, and their corresponding graphs are presented in the lower part of the diagram.}
    \label{fig:1}
\end{figure}

Luque et al. in~\cite{luque2009horizontal} presented a more straightforward technique called horizontal visibility graph (HVG) in 2009 to decrease the computational cost associated with NVG. According to this method, if it is feasible to draw a horizontal line in the time series linking the two vertical bars corresponding to the two data that do not intercept any height of the intermediate data in between, two nodes in the graph are connected in this option. We show a basic demonstration of this algorithm in the middle part of Figure~\ref{fig:1}, using a toy time series and the resultant horizontal visibility graph; the orange lines indicate the horizontal visibility lines between all of the observations, and each data in the sequence corresponds to a graph node, and two nodes are connected if their corresponding data heights are greater than the sum of all data heights between them. 
Formally, If the following criterion is met, two nodes $(t_a,y_a)$ and $(t_b,y_b)$  are linked and are horizontally visible. 
\begin{equation}  y_a,y_b>y_c   \end{equation}
For all $t_c$ such that $t_a<t_c<t_b$.

\subsubsection{Quantile graph}
Campanaro et al in~\cite{campanharo2011duality} presented quantile graph (QG) in 2011. This method is distinct from prior visibility approaches, it collects relevant time series information such as oscillations across time. The time series is divided into Q quantiles, $q_1,q_2,\cdots,q_Q,$ which each quantile, $q_i$, represents a network node $v_i$ in the network. As a result, the graph has the same number of nodes as the number of quantiles. A weighted directed link $(v_a, v_b, w_{(a,b)})$ connects two vertices $v_a$ and $v_b$, where the weight $w_{(a,b)}$ reflects the amount of times an observation $(t_n, y_n)$ from the quantile $q_a$ is followed by an observation $(t_{(n+1)}, y_{(n+1)})$ from the quantile $q_b$.

Formally, let $M$ be a transition of the time series $X\in T$ to a graph $g\in G$, with~$X = \{x(t) \vert t \in N,x(t) \in R\}$ and $g= {V,E}$ is a collection of nodes (vertices) $V$ and edges $E$, respectively. Assumes a basic discretization of $X$ that is not affected by its value distribution. Ranking $X$ and breaking this ranking across the $Q-1$ intervals identify the $Q$ quantiles of a time series, defined by the cutting points that divide $X$ into $Q-1$ equally sized intervals~\cite{zar1999biostatistical}. Each quantile is defined as follows:
\begin{equation} q_i = x(\frac{T*i}{Q})   \end{equation}
Consider $i = 1,2,\cdots,Q$ each $X$ value is mapped into the associated $q_i$ For $i = 1,2,\cdots,Q$ once the quantiles of $Q$ have been established. Then the node $v_i  \in V$ of the relevant network is assigned to each quantile $q_i$. In a network with a weighted arc $(v_i,v_j,w_{(i,j)}^k) \in E$ two nodes $v_i$ and $v_j$ are linked by $E$ when there exist two values $x(t)$ and $x(t+1)$ of $q_i$ and $q_j$, of $t = 1,2,\cdots,T$, where $k$ is time differences $k = 1,\cdots,k_{Max}$, in this study we consider a case when the time difference equal to one.
Each weight $w_{(i,j)}^k$  in the weighted directed adjacency matrix, which is denoted as $A_K$, is the number of times a value in the quantile has occurred $q_i$ at time $t$ is followed by a point in quantile $q_j$ At the time $t+k$. As a result, successive transitions along the same arc raise the related weight's value. The weighted directed adjacency matrix $A_K$ forms a Markov transition matrix $W_K$  after adequate normalization, with $\sum w_{(i,j)}^k=1$~\cite{campanharo2016hurst}. In the right side of Figure~\ref{fig:1}, we depict this projection and the associated network using a toy time series; it is illustrated from a uniform white noise $X$ with $T = 20$ time points, $Q = 4$ quantiles. In the time-series picture, the various colors reflect the region relating to the different quantiles. Repeated transitions among quantiles lead to graph edges with greater weights indicated by thicker lines.

\subsection{Graph isomorphism network}
In this paper, for the graph classification task, we utilize a branch of graph neural network named graph isomorphism network (GIN)~\cite{xu2018powerful}; graph isomorphism network is inspired by the tight association between the Weisfeiler-Lehman test~\cite{weisfeiler1968reduction} and graph neural networks, Weisfeiler-Lehman test is a potent test known to differentiate a large class of graphs~\cite{babai1979canonical}. Consider  $G = (V, E)$ as a graph, with node feature vectors $X_v$ For $v \in V$. Node classification and graph classification, are two components that are of particular interest in~\cite{xu2018powerful}, in this study we applied graph classification task, in which given a set of graphs $\{G_1,\cdots,G_N \}\subseteq g$ and their labels $\{y_1,\cdots, y_N \} \subseteq y$ the aim is to learn a representation $h_G$, it assists in the prediction of an entire graph's label $y_G= f(h_G )$. Graph neural networks learn a representation vector of a node $h_v$ and $h_G$, which is the representation vector of the graph as a whole based on the graph topology and node characteristics $X_v$. Modern GNNs use a neighborhood aggregation approach, which updates the node's representation iterative by aggregating its neighbors' representations; after $k$ iterations of aggregation, the structural information throughout a node's $k-hop$ network neighborhood is included in its representation. by making $\epsilon$ as a learn-able parameter or a fixed scalar, the node representations are updated by GIN in the following way: 

\begin{equation} h_v^{(k)} = MLP^{k}\left((1+\epsilon^{k}).h_v^{(k-1)}+\sum_{u\in{N(v)}}h_v^{(k-1)}\right) \end{equation}
Which $h_v^{(k)}$ represents the node $v$ feature vector in the $k-th$ iteration/layer. We assign $h_v^{(0)}=X_v$  as the starting value and $N(v)$ is a group of nodes that are adjacent to $v$.
GIN's node embeddings could be utilized directly for tasks such as node classification and link prediction and graph classification. Researchers~\cite{xu2018powerful} proposed the following readout function for graph classification tasks, which creates the embedding of the whole graph from individual node embeddings:

\begin{equation} h_G = C\left(R\left(\{h_v^{\left(k\right)} \vert v\in G \}\right), k=0, 1, \cdots, K\right) \end{equation}
Where $C$ is concatenating function, and $R$ is a readout function.
\section{Experiment}

\begin{figure}[H]
    \begin{center}
       \includegraphics[width = 1\textwidth]{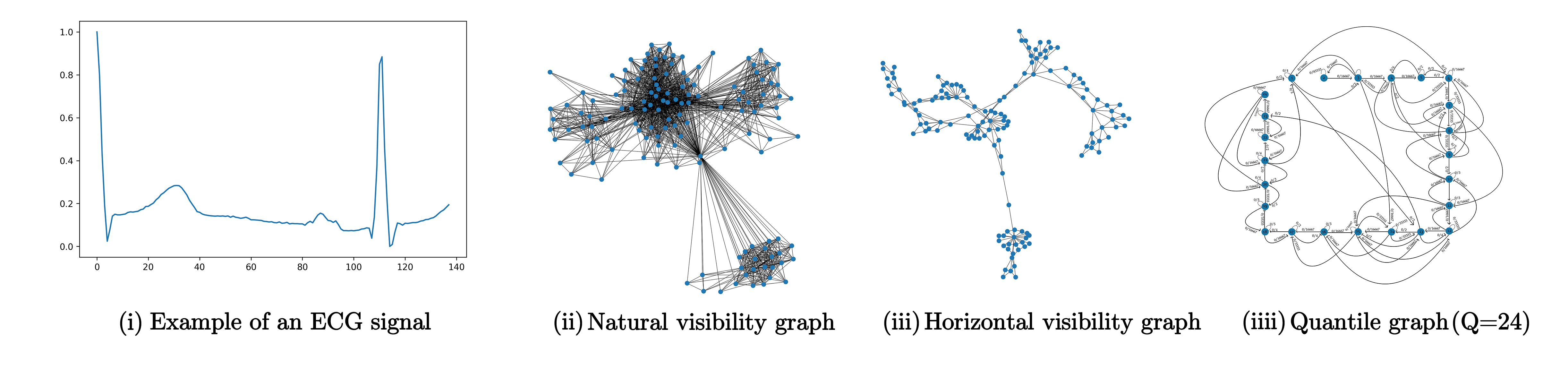} 
    \end{center}
    \caption{(i) An example of a normal ECG signal, and its corresponding (ii) natural visibility graph, (iii) horizontal visibility graph, and (iiii) quantile graph when the number of quantiles is equal to 24.}
    \label{fig:2}
\end{figure}

We use the PTB diagnostic ECG database as a data source for labeled ECG recordings in this study~\cite{goldberger2000physiobank} and~\cite{bousseljot1995nutzung}. We used ECG lead II re-sampled to a sampling frequency of 125Hz as the input in all our investigations. The PTB Diagnostics dataset contains ECG records from 290 people, with 148 being diagnosed with MI (Myocardial Infarction), 52 being healthy samples, and the remainder having seven other diseases. ECG signals from 12 leads were captured at a frequency of 1000Hz in each record. In this study, we used preprocessed PTB diagnostic ECG data~\cite{kachuee2018ecg}. This paper proposes a straightforward yet efficient preprocessing approach for ECG signals, which involves using input ECG beats. The steps for extracting beats from an ECG signal are as follows~\cite{kachuee2018ecg}:
\begin{enumerate}
    \item Segment the continuous ECG signal into 10s windows, select one 10s window.
    \item Normalize the amplitude values within the range of 0-1.
    \item  Locate the collection of local maxima with the help of the zero-crossings of the first derivative.
    \item Find the set of possible R-peak candidates using a threshold of 0.9 on the normalized value of the local maximums.
    \item Calculate the median of R-R time intervals as the nominal heartbeat period of the window (T).
    \item For each R-peak, select a signal part with a length of 1.2T.
    \item Fill each chosen part with zeros to a predefined fixed length.
\end{enumerate}
It is noteworthy that the proposed beat extraction method is highly successful in extracting R-R intervals from signals with varied morphology's. Moreover, it is also capable of accurately recognizing changes in heart rate, which is a crucial step for ECG signal analysis.
Regarding the number of the abnormal samples, which is considerably more than the normal samples, we duplicate the normal samples by simply copying and pasting the normal data to tackle the propensity of the graph neural network to predict the larger sample category.
The first task in the methodology suggested in the paper was transforming ECG signals into complex networks. We created three distinct types of complex networks for the ECG signals under investigation. The first approach was Natural Visibility Graph (NVG)~\cite{lacasa2008time}, the second method was horizontal visibility graph (HVG)~\cite{luque2009horizontal} and quantile graph was the third method that we used to convert the ECG signal to the graph (QG)~\cite{campanharo2011duality}, furthermore to applying quantile graph approach we had to choose the number of quantiles for this step, so we used the whole data set to transform the time series into a quantile graph using different quantiles from $2$ to $30$ and fed the graphs to the graph isomorphism network, then the quantile number with the best performances picked, which was $q=24$ in our study. An example of an ECG signal and its corresponding natural visibility graph, horizontal visibility graph, and quantile graph when the number of quantiles is equal to 24  is depicted in Figure~\ref{fig:2}. In the second phase, we tried to train a model to classify time series corresponding graphs, and we fed graphs of converted time series to a graph neural network named graph isomorphism network to develop a model for classifying time series corresponding graphs.

\begin{figure}[H]
    \begin{center}
       \includegraphics[width = 0.58\textwidth]{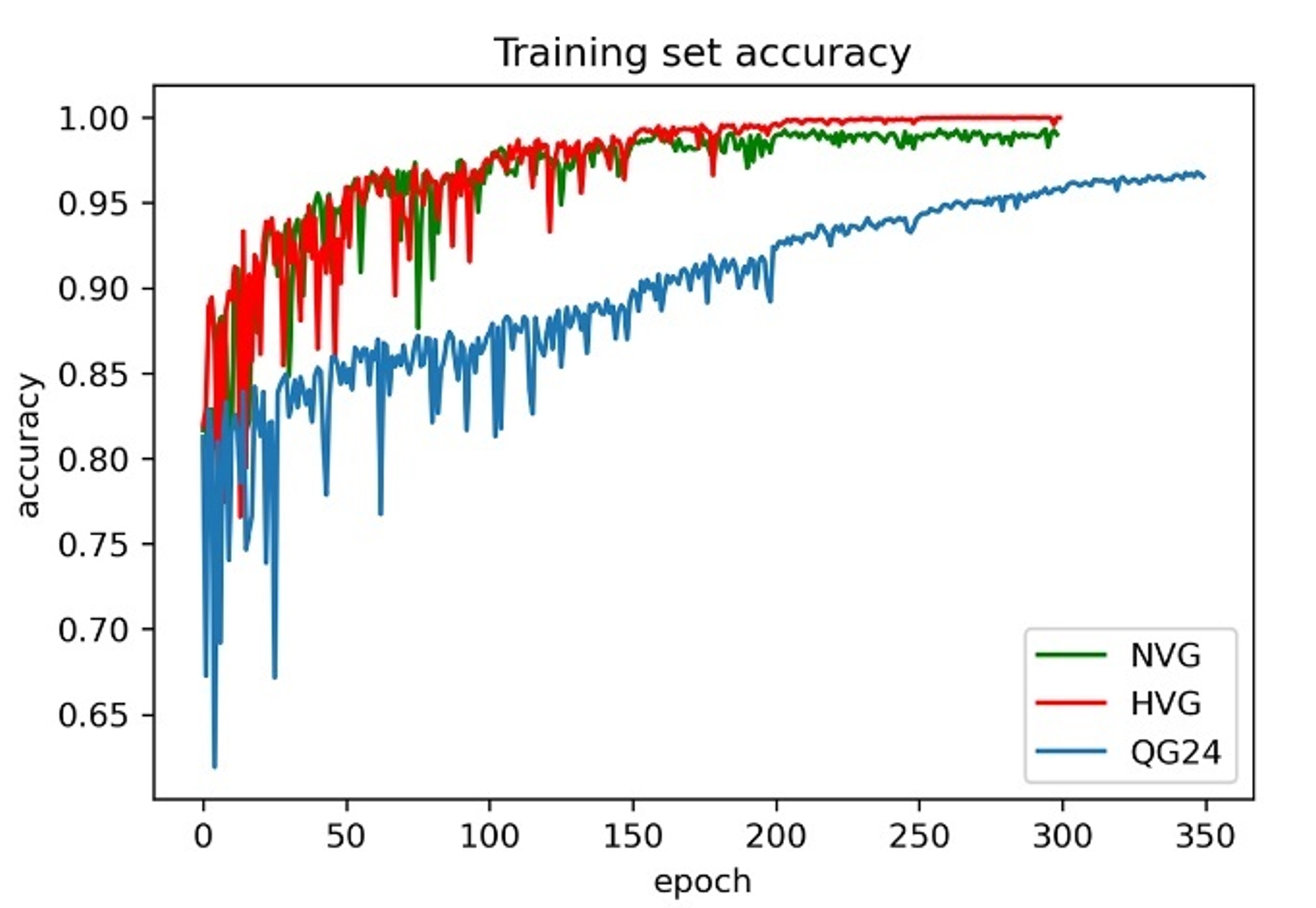} 
    \end{center}
    \caption{Training set performance associated with each proposed method. We can see that both NVG and HVG have better performance than QG-24 in the training set, NVG and HVG are able to almost perfectly fit the training set. }
    \label{fig:3}
\end{figure}

All the experiments are carried out in Python; for converting time series to graphs using the three mentioned methodologies, we developed an application in Python. We utilize the official graph isomorphism network (GIN) software~\cite{xu2018powerful} that authors of~\cite{xu2018powerful} published to classify the graphs.
We assess the graph isomorphism network (GIN) (Equations 4 and 5) for the model and configuration for the graph classification task. Following ~\cite{niepert2016learning},~\cite{yanardag2015deep} and~\cite{xu2018powerful}, we implement LIB-SVM to conduct 10-fold cross-validation. Within the cross-validation, we present the average validation accuracies throughout the 10-fold. GNN layers between five and eight (consisting of the input layer) are applied across all setups, and all MLPs have two layers. Every hidden layer goes through batch normalization~\cite{ioffe2015batch}, and the adam optimizer~\cite{kingma2014adam} has been used, with a starting learning rate of 0.01, and every 50 epochs, the learning rate declines by 0.5. The hyper-parameters are tuned as follows, the number of the hidden neurons was 64, batch size was between 32, 64, and 128, and dropout ratio after the dense layer was between 0 and 0.5~\cite{srivastava2014dropout}, and the number of epochs was between 300 and 400.
The epoch with the maximum cross-validation accuracy averaged over the ten folds, was chosen.
In Figure~\ref{fig:3} we plot the training set performance of training using different transformation methods that we used in this study.
Table~\ref{tab:1} compares the experiment described in this paper with other approaches in the literature that solve the same classification problem using the same data set we used in our investigation.

\begin{table}[h]
\begin{center}
\caption{Comparison of the proposed methods with other methods in the literature.}\label{tab:1}%
\begin{tabular}{@{}lll@{}}
\toprule 
Rank & Method & Accuracy\%\tablefootnote{Average of ten-fold cross-validation.} \\
\midrule
 {\bfseries 1} & {\bfseries Proposed approach using natural visibility graph.} & {\bfseries99.38\%}\\
$2$ & Ahmad, Tabassum, Guan \& Khan, 2021~\cite{ahmad2021ecg} & 99.2\% \\

$3$ & El Boujnouni, Imane, et PTB, 2022~\cite{el2022wavelet} & 98.8\% \\

 {\bfseries 4} & {\bfseries Proposed approach using horizontal visibility graph.} & {\bfseries98.76\% }\\
$5$ & Ahamed, Hasan, Monowar, Mashnoor \& Hossain, 2020~\cite{ahamed2020ecg} & 97.66\% \\
$6$ & Chen, Chen, He, Yang \& Cao, 2018 .~\cite{chen2018multi} & 96.18\% \\
$7$ & Liu et al., 2017 ~\cite{liu2017real} & 96\% \\
$8$ & Sharma, Tripathy \& Dandapat, 2015~\cite{sharma2015multiscale} & 96\% \\
$9$ &  Kachuee, Fazeli \& Sarrafzadeh, 2018~\cite{kachuee2018ecg} & 95.9\% \\
$10$ &  Kojuri, Boostani, Dehghani, Nowroozipour\& Saki, 2015~\cite{kojuri2015prediction} & 95.9\% \\
$11$ &  Kumar, Sanjay, et al. 2023~\cite{kumar2023fuzz} & 95.79\% \\

$12$ & Acharya et al., 2017~\cite{acharya2017application} & 95.22\% \\
 {\bfseries 13}&   {\bfseries Proposed approach using quantile graph with 24 quantiles.} & {\bfseries91.93\%} \\
$14$ &  Diker, C{\"o}mert, Avc{\i}, To{\u{g}}a{\c{c}}ar\& Ergen, 2019~\cite{diker2019novel} & 83.82\% \\
\botrule
\end{tabular}
\end{center}
\end{table}

As we can see, the proposed approach using the NVG graph with 99.38\% accuracy has the highest performance, followed by the Proposed method using the HVG graph with 98.76\% accuracy, and the proposed approach using QG-24 with 91.93\% accuracy is the third one. Consequently, we may conclude that this method is suitable for classifying ECG signals.
%------------------------------------------------

\section{Conclusion}
The suggestion of integrating mapping approaches using different NVG, HVG, and QG to create a single richer set of features defining a time series for classifying the time series using associated complex networks by implementing a graph neural network named graph isomorphism network was our main contribution. We may conclude from the results obtained in this paper that these various mapping techniques can be used effectively for classifying ECG signals using a graph isomorphism network. This is the first time such a comprehensive and methodical investigation has been conducted. By using the graph isomorphism network in our experiment, we can deduce from the findings obtained in this paper that this mapping method can be appropriately used for ECG signal classification. In this paper, we attempted to map the time series to the natural visibility graph, horizontal visibility graph, and quantile graph, then feed the graphs to the graph isomorphism network in order to classify the ECG signal. We achieved 99.38\% accuracy for the first proposed approach using the NVG graph, for the second proposed approach using the HVG graph, we got 98.76\% accuracy, and for the third proposed method QG-24, we got 91.93\% accuracy.
Future studies can investigate other GNNs' performance on ECG signal categorization. 
The current univariate mappings can be generalized to multivariate settings. For example, we can investigate multilayer complex networks, which allow multiple distinct variables (for example, ECG and EEG signals may be represented on a separate layer in the same network) and spatial characteristics to be combined in a single complicated network; after that, we can use graph neural networks for classifying combination of two or more time series.
On the other hand, the proposed approach can apply to other tasks in time series classification, such as voice recognition.

\section{Acknowledgments}
This work was also supported by TAILOR, a project funded by the EU Horizon 2020 research and innovation program under GA No 952215.


%\bibliography{sn-bibliography}

\begin{thebibliography}{31}

\bibitem{acharya2017application} Acharya, U.R., Fujita, H., Oh, S.L., Hagiwara, Y., Tan, J.H., Adam, M.: Application of deep convolutional neural network for automated detection of myocardial infarction using ecg signals. Information Sciences 415, 190–198 (2017)

\bibitem{ahamed2020ecg} Ahamed, M.A., Hasan, K.A., Monowar, K.F., Mashnoor, N., Hossain, M.A.: Ecg heartbeat classification using ensemble of efficient machine learning approaches on imbalanced datasets. In: 2020 2nd International Conference on Advanced Information and Communication Technology (ICAICT). pp. 140–145. IEEE (2020)

\bibitem{ahmad2021ecg} Ahmad, Z., Tabassum, A., Guan, L., Khan, N.: Ecg heart-beat classification using multimodal image fusion. In: ICASSP 2021-2021 IEEE International Conference on Acoustics, Speech and Signal Processing (ICASSP). pp. 1330–1334. IEEE (2021)

\bibitem{al2016deep} Al Rahhal, M.M., Bazi, Y., AlHichri, H., Alajlan, N., Melgani, F., Yager, R.R.:
Deep learning approach for active classification of electrocardiogram signals. Information Sciences 345, 340–354 (2016)

\bibitem{babai1979canonical} Babai, L., Kucera, L.: Canonical labelling of graphs in linear average time. In: 20th
Annual Symposium on Foundations of Computer Science (sfcs 1979). pp. 39–46. IEEE (1979)

\bibitem{bousseljot1995nutzung} Bousseljot, R., Kreiseler, D., Schnabel, A.: Nutzung der EKG-Signaldatenbank CARDIODAT der PTB {\"u}ber das Internett (1995)

\bibitem{campanharo2016hurst} Campanharo, A.S., Ramos, F.M.: Hurst exponent estimation of self-affine time
series using quantile graphs. Physica A: Statistical Mechanics and its Applications 444, 43–48 (2016)

\bibitem{campanharo2011duality} Campanharo, A.S., Sirer, M.I., Malmgren, R.D., Ramos, F.M., Amaral, L.A.N.:
Duality between time series and networks. PloS one 6(8), e23378 (2011)

\bibitem{chen2018multi} Chen, Y., Chen, H., He, Z., Yang, C., Cao, Y.: Multi-channel lightweight convolution neural network for anterior myocardial infarction detection. In: 2018 IEEE SmartWorld, Ubiquitous Intelligence \& Computing, Advanced \& Trusted Computing, Scalable Computing \& Communications, Cloud \& Big Data Computing, Internet of People and Smart City Innovation (Smart-World/SCALCOM/UIC/ATC/CBDCom/IOP/SCI). pp. 572–578. IEEE (2018)

\bibitem{diker2019novel} Diker, A., C{\"o}mert, Z., Avc{\i}, E., To{\u{g}}a{\c{c}}ar, M., Ergen, B.: A novel application based
on spectrogram and convolutional neural network for ecg classification. In: 2019 1st International Informatics and Software Engineering Conference (UBMYK). pp. 1–6. IEEE (2019)

\bibitem{goldberger2000physiobank} Goldberger, A.L., Amaral, L.A., Glass, L., Hausdorff, J.M., Ivanov, P.C., Mark,
R.G., Mietus, J.E., Moody, G.B., Peng, C.K., Stanley, H.E.: Physiobank, physiotoolkit, and physionet: components of a new research resource for complex physiologic signals. circulation 101(23), e215–e220 (2000)

\bibitem{hammad2019novel} Hammad, M., Zhang, S., Wang, K.: A novel two-dimensional ecg feature extraction and classification algorithm based on convolution neural network for human authentication. Future Generation Computer Systems 101, 180–196 (2019)

\bibitem{ioffe2015batch} offe, S., Szegedy, C.: Batch normalization: Accelerating deep network training by reducing internal covariate shift. In: International conference on machine learning. pp. 448–456. PMLR (2015)

\bibitem{kachuee2018ecg} Kachuee, M., Fazeli, S., Sarrafzadeh, M.: Ecg heartbeat classification: A deep transferable representation. In: 2018 IEEE international conference on healthcare informatics (ICHI). pp. 443–444. IEEE (2018)

\bibitem{kingma2014adam}  Kingma, D.P., Ba, J.: Adam: A method for stochastic optimization. arXiv preprint
arXiv:1412.6980 (2014)

\bibitem{kojuri2015prediction} Kojuri, J., Boostani, R., Dehghani, P., Nowroozipour, F., Saki, N.: Prediction
of acute myocardial infarction with artificial neural networks in patients with nondiagnostic electrocardiogram. Journal of Cardiovascular Disease Research 6(2) (2015)

\bibitem{lacasa2008time} Lacasa, L., Luque, B., Ballesteros, F., Luque, J., Nuno, J.C.: From time series to complex networks: The visibility graph. Proceedings of the National Academy of Sciences 105(13), 4972–4975 (2008)

\bibitem{liu2017real} Liu, W., Zhang, M., Zhang, Y., Liao, Y., Huang, Q., Chang, S., Wang, H., He, J.: Real-time multilead convolutional neural network for myocardial infarction detection. IEEE journal of biomedical and health informatics 22(5), 1434–1444 (2017)

\bibitem{luque2009horizontal} Luque, B., Lacasa, L., Ballesteros, F., Luque, J.: Horizontal visibility graphs: Exact
results for random time series. Physical Review E 80(4), 046103 (2009)

\bibitem{marinho2019novel} Marinho, L.B., de MM Nascimento, N., Souza, J.W.M., Gurgel, M.V., Rebou{\c{c}}as Filho, P.P., de Albuquerque, V.H.C.: A novel electrocardiogram feature extraction approach for cardiac arrhythmia classification. Future Generation Computer Systems 97, 564–577 (2019)

\bibitem{mondejar2019heartbeat}  Mond{\'e}jar-Guerraa, V., Novo, J., Rouco, J., Penedo, M.G., Ortega, M.: Heartbeat classification fusing temporal and morphological information of ecgs via ensemble of classifiers. Biomedical Signal Processing and Control 47, 41–48 (2019)

\bibitem{niepert2016learning} Niepert, M., Ahmed, M., Kutzkov, K.: Learning convolutional neural networks for graphs. In: International conference on machine learning. pp. 2014–2023. PMLR (2016)

\bibitem{sharma2015multiscale} Sharma, L., Tripathy, R., Dandapat, S.: Multiscale energy and eigenspace approach to detection and localization of myocardial infarction. IEEE transactions on biomedical engineering 62(7), 1827–1837 (2015)

\bibitem{sharma2018novel} Sharma, M., San Tan, R., Acharya, U.R.: A novel automated diagnostic system for classification of myocardial infarction ecg signals using an optimal biorthogonal filter bank. Computers in biology and medicine 102, 341–356 (2018)

\bibitem{srivastava2014dropout} Srivastava, N., Hinton, G., Krizhevsky, A., Sutskever, I., Salakhutdinov, R.: Dropout: a simple way to prevent neural networks from overfitting. The journal of machine learning research 15(1), 1929–1958 (2014)

\bibitem{utomo2019automatic} Utomo, O.K., Surantha, N., Isa, S.M., Soewito, B.: Automatic sleep stage classification using weighted elm and pso on imbalanced data from single lead ecg. Procedia Computer Science 157, 321–328 (2019)

\bibitem{weisfeiler1968reduction} Weisfeiler, B., Leman, A.: The reduction of a graph to canonical form and the
algebra which appears therein. NTI, Series 2(9), 12–16 (1968)

\bibitem{xu2018powerful} Xu, K., Hu, W., Leskovec, J., Jegelka, S.: How powerful are graph neural networks?
arXiv preprint arXiv:1810.00826 (2018)

\bibitem{yanardag2015deep} Yanardag, P., Vishwanathan, S.: Deep graph kernels. In: Proceedings of the 21th ACM SIGKDD international conference on knowledge discovery and data mining. pp. 1365–1374 (2015)

\bibitem{yeh2010novel} Yeh, Y.C., Wang, W.J., Chiou, C.W.: A novel fuzzy c-means method for classifying heartbeat cases from ecg signals. Measurement 43(10), 1542–1555 (2010)

\bibitem{zar1999biostatistical} Zar, J.H.: Biostatistical analysis. Pearson Education India (1999)

\bibitem{el2022wavelet}El Boujnouni, I., Zili, H., Tali, A., Tali, T. \& Laaziz, Y. A wavelet-based capsule neural network for ECG biometric identification. {\em Biomedical Signal Processing And Control}. \textbf{76} pp. 103692 (2022)

\bibitem{kumar2023fuzz}Kumar, S., Mallik, A., Kumar, A., Del Ser, J. \& Yang, G. Fuzz-ClustNet: Coupled fuzzy clustering and deep neural networks for Arrhythmia detection from ECG signals. {\em Computers In Biology And Medicine}. pp. 106511 (2023)



\end{thebibliography}

\end{document}